\documentclass[fleqn,10pt]{wlscirep}%
\usepackage{microtype}

\usepackage{breakurl}

\usepackage{amsmath}%
\usepackage{amsfonts}%
\usepackage{amssymb}%
\usepackage{amsthm}
\usepackage{graphicx}
\usepackage[utf8]{inputenc} 

\usepackage{bm}
\usepackage{authblk}
\usepackage{cite}

\newcommand{\dmud}{{\Delta\bar\mu_{\uparrow\downarrow}}} 
\newcommand{\Js}{\vec{J_{\mathrm{s}}}}

\renewcommand{\vec}[1]{\boldsymbol{#1}}  
\newcommand{\muB}{$\mu_{\mathrm{B}}$}

\title{A Microscopic Explanation of Microwave Spin Pumping in Spintronics}
\author[1,*]{John F. Gregg}
\author[2]{Burkard Hillebrands}
\author[1]{Alistair Inglis}
\author[1]{Monika E. Mycroft}
\author[2]{Evangelos Th. Papaioannou}
\author[1]{James Semple}
\author[1]{Calvin J. Tock}

\affil[1]{Department of Physics, University of Oxford, Clarendon Laboratory, Parks Road, Oxford OX1 3PU, UK}
\affil[2]{Fachbereich Physik, Technische Universit\"at Kaiserslautern, 67663 Kaiserslautern, Germany}

\affil[*]{john.gregg@magd.ox.ac.uk}

\begin{abstract}
We contend that Microwave Spin Pumping was first predicted and observed - albeit using a different and more sensitive detection mechanism than Inverse Spin Hall Effect - in the 1950’s. This discovery was the founding step in the widely used analytical tool that is now known as Dynamic Nuclear Polarisation. Recognising this hitherto unsung connection between $20^{\textrm{th}}$ Century Magnetic Resonance and $21^{\textrm{st}}$ Century Spintronics not only helps to explain and unify contemporary metallic spin pumping observations: it is also the key to unlocking the immense and very sophisticated toolbox of Magnetic Resonance and placing it at the disposal of the future of Spintronics.

\end{abstract}

\begin{document}

\flushbottom
\maketitle

Microwave spin pumping in metallic spintronic systems has recently been widely reported\cite{doi:10.1063/1.4871514,PhysRevB.89.235317,doi:10.1063/1.4792693,doi:10.1063/1.3587173,doi:10.1063/1.4918909,PhysRevB.82.214403,Ando2011,PhysRevApplied.8.014022,PhysRevB.95.064406,7927441,PhysRevB.93.134405,Papaioannou2013,PhysRevB.95.174426,PhysRevB.96.024437} and the effect has been empirically and phenomenologically justified \cite{PhysRevB.82.214403}. We aim here to provide a microscopic physical explanation for the process, futhering the insights provided in \cite{PhysRevB.96.064423,Watts2006}. In this Letter we present a basic quantum mechanical explanation for metallic Microwave Spin Pumping. In doing so, we draw on the physics of a well-known and comprehensively understood technique from an apparently completely different branch of condensed matter physics: Dynamic Nuclear Polarisation (DNP) \cite{abragam1982nuclear}. We argue that the electronic aspects of certain versions of DNP are identical to what happens in metallic Spintronic Microwave Spin Pumping.

In the early days of magnetic resonance, Albert Overhauser made a shocking prediction\cite{Overhauser1953} that appeared to fly in the face of Thermodynamics\cite{slichter1996principles}: namely that in a non-magnetic metal that is subject to an applied magnetic field, the nuclear spins could be polarised by strong microwave irradiation of the itinerant electronic magnetic resonance transition\cite{abragam1961principles}. Despite the widespread scepticism with which this proposition was initially received in the magnetic resonance community, Overhauser's prediction was experimentally vindicated soon afterwards in a seminal experiment by Carver and Slichter\cite{Carver1956}, before becoming generally accepted\cite{Gueron1959,Hecht1963}. This marked the birth of the family of techniques that are now known under the umbrella term Dynamic Nuclear Polarisation (DNP).

The Overhauser Effect can be understood as a two-step process: the first step uses microwave irradiation to create a non-equilibrium electronic state: the second step is a cross-relaxation process - mediated by the scalar coupling between electrons and nuclei - by virtue of which the non-equilibrium electronic state polarises the nuclei. For the avoidance of confusion, it is the creation and nature of the initial non-equilibrium electronic state that is of primary interest in the context of Spintronic Microwave Spin Pumping; but we will return to the significance and experimental potential of the second step at the end of this Letter.

\begin{figure} 
	\centering
		\includegraphics[width=1.00\textwidth]{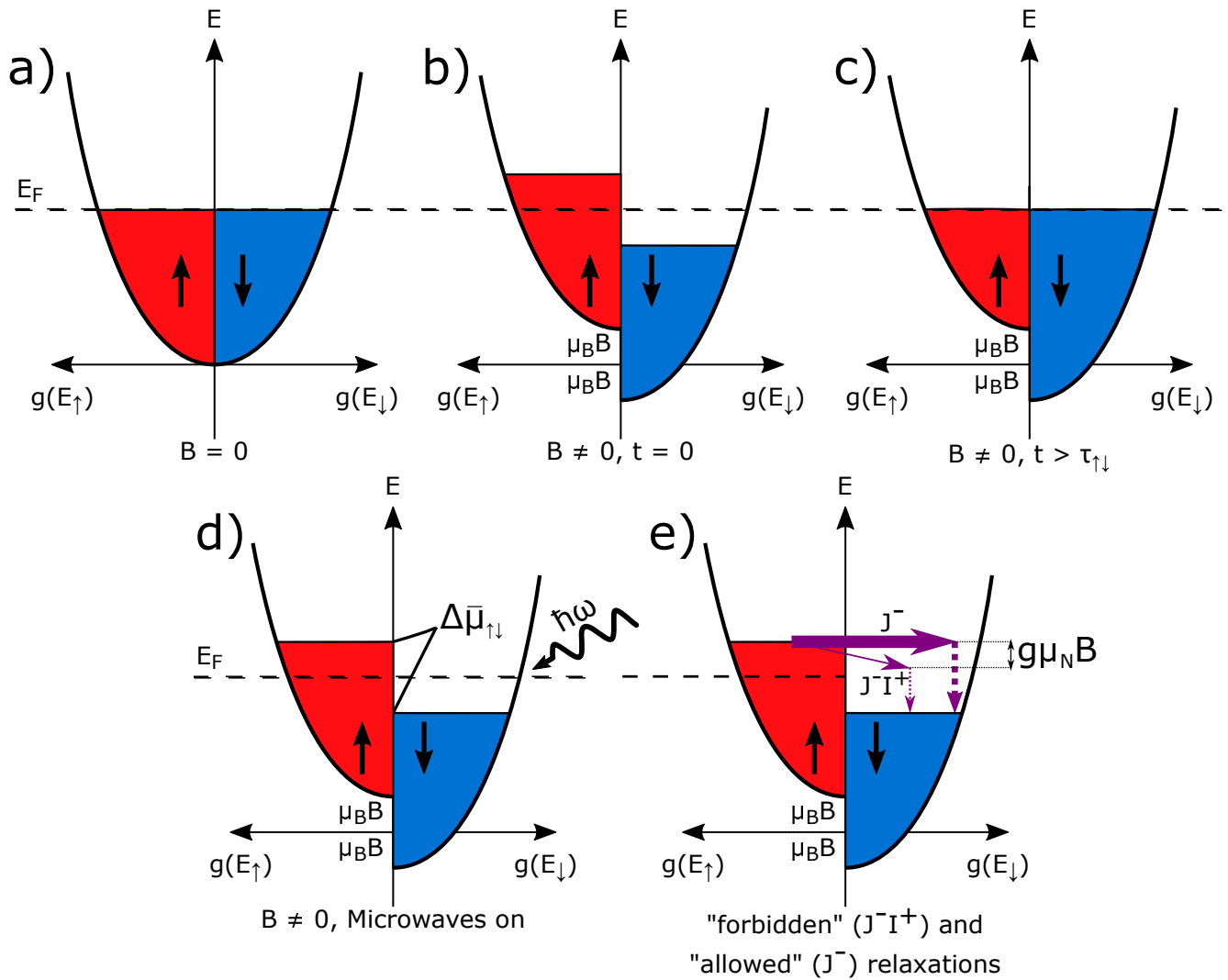}
	\caption{\textbf{Bandstructure of a simple paramagnetic metal} Figure \ref{fig:PauliDiagram}a shows the idealised bandstructure of a simple paramagnetic metal in zero applied magnetic field. Figure \ref{fig:PauliDiagram}b shows the bandstructure immediately following the instantaneous application of magnetic field: note the transient appearance of an electrochemical potential split between the spin channels. The subsequent spin relaxation causes evolution to the state shown in Figure \ref{fig:PauliDiagram}c. In Figure \ref{fig:PauliDiagram}d an electrochemical potential splitting or spin accumulation has reappeared following microwave irradiation. Figure \ref{fig:PauliDiagram}e shows the ``allowed''  $J^{-}$ and ``forbidden'' $J^{-}I^{+}$ - relaxation processes that tend to erode the microwave-induced spin accumulation; the ``forbidden'' events are capable of polarizing the nuclei.}
	\label{fig:PauliDiagram}
\end{figure}

Figure \ref{fig:PauliDiagram} explains the operation of the Overhauser Effect. Figure \ref{fig:PauliDiagram}a shows the idealised electronic bandstructure of a simple non-magnetic metal in zero applied magnetic field. On instantaneous application of a magnetic field, the picture changes to that shown in Figure \ref{fig:PauliDiagram}b where the electrochemical potentials of the two electronic spin channels are different by an energy splitting of $\Delta E = 2$\muB${B}$; this causes the onset of spin relaxation that equalises the electrochemical potentials on a timescale $\tau_{\uparrow\downarrow}$ to produce the diagram of Figure \ref{fig:PauliDiagram}c.

The metal of Figure \ref{fig:PauliDiagram}c is now subjected to microwave irradiation, which works to equalize the populations of the two spin bands\cite{Foot2005} and causes a net transfer of electron spins from the lower lying, more populated spin band to its upper counterpart. This causes the reappearance of an electrochemical potential splitting between the spin channels whose magnitude is the result of a dynamic equilibrium between the microwave-induced spin transition rate and spin-flip relaxation back in the reverse direction as shown in Figure \ref{fig:PauliDiagram}d. In the limit of a very strong microwave field, the population imbalance between the two spin bands is removed and the full electrochemical potential splitting of $\dmud = 2$\muB${B}$ is recovered. In the parlance of Spintronics, this non-equilibrium electrochemical splitting of the spin channels is nothing other than a \textbf{spin accumulation}, and it is this feature of the Overhauser Effect - whose appearance is apparent even in the diagrams of Overhauser's original paper\cite{Overhauser1953} - that we argue is capable of generating charge-decoupled pure electronic spin currents.

Figure \ref{fig:PauliDiagram}e explains the second step of the Overhauser Effect (which is of no immediate interest to us in this Letter but which sets the scene for a point we discuss later, namely that Nuclear Magnetic Resonance (NMR) is potentially a more sensitive experimental probe of spin accumulations than is Inverse Spin Hall Effect (ISHE) ) and shows the relaxation processes that work to destroy the spin accumulation. There are two such types of relaxation: processes that involve only electron spin-flips and less frequent processes involving mutual electron-nuclear spin flip-flops that proceed by virtue of the scalar form of the electron-nuclear coupling Hamiltonian; it is these latter transitions that drive the second stage of the Overhauser Effect and generate a nuclear spin polarisation.

Our dissection of the Overhauser Effect into two distinct stages - firstly the creation of a spin accumulation and secondly the action of spin-lattice relaxation which polarises the nuclei - leads us to predict that any other method of creating a spin accumulation in the same material will likewise lead to nuclear polarisation. For example, tunnel-injecting a spin polarised current into a semiconductor such as Gallium Arsenide should lead to formation of a spin accumulation in the semiconductor and hence to polarisation of the Ga nuclei via electron-nuclear scalar coupling. It turns out that exactly this effect has already been observed \cite{Shiogai2012}, thereby lending further weight to our interpretation of the underlying science of the Overhauser Effect.

Returning to the microwave Overhauser mechanism that led to the spin accumulation of Figure \ref{fig:PauliDiagram}d, we now consider what happens if - instead of a paramagnetic metal - we choose a ferromagnetic metal with exchange-split spin sub-bands as the Overhauser host material. Again we apply an external magnetic field: but this time we microwave irradiate at the ferromagnetic resonance frequency which is determined by the applied magnetic field and the demagnetising factor of the specimen. The response is a precessing magnetic moment represented by a coherent macroscopic quantum state comprising a large number of single particle states\cite{kittel2004introduction}. For modest microwave power, the precession angle is small, so the exchange field responsible for the magnetization $\vec{M}$ and the applied magnetic field almost agree as to the quantisation axis. The single particle states that subscribe to the macroscopic precession of the $\vec{M}$ vector therefore contain a small admixture of the spin-down, anti-parallel aligned state with respect to the applied magnetic field quantization axis. Longitudinal spin relaxation events - which are represented in classical language by the Landau-Lifshitz-Gilbert Damping Factor\cite{stancil2009spin} - collapse single particle wavefunctions, a fraction of which emerge spin-down. In the coherent quantum state treatment of cooperative magnetism, these spin-flips can be thought of as magnonic excitations\cite{ashcroft1976solid}: but in the single-particle description that is the accepted spintronic treatment of spin injection from paramagnets to ferromagnets as introduced by Valet and Fert\cite{Valet1993}, we recover the picture of two spin channels with different electrochemical potentials. In other words, we have again arrived at a spin accumulation: but with an important difference. The asymptotic limit to the electrochemical potential splitting in the limit of large microwave radiation is not now the microwave quantum (that is determined by the ferromagnetic resonance frequency and hence by the magnetic field applied); it is now the very much larger exchange splitting of the ferromagnet. There is thus a clear advantage to implementing Overhauser spin pumping in a ferromagnetic metal: typical exchange splittings in such metals are equivalent to electronic Zeeman splittings in magnetic fields that are virtually unachievable in the laboratory; therefore a much larger spin accumulation may be achieved by Overhauser pumping in an ordered magnetic metal than in a paramagnet. Nevertheless, we maintain that the Overhauser spin pumping principle is equally applicable to both.

This conclusion is robustly supported by the recent work of Tatara and Mizukami\cite{PhysRevB.96.064423}. Although these authors do not explicitly address the Overhauser Effect, equations 8-15 of their paper constitute a robust formal proof that Overhauser pumping of ferromagnetic metals results in a spin accumulation whose asymptote is the exchange quantum as opposed to the electronic Zeeman quantum. It is also interesting to note that, suitably modified, the same section of their analysis may be ``reverse engineered'' to rediscover by an alternative and more formal route the original prediction for paramagnetic metals at which Overhauser arrived from simple physical considerations.

\begin{figure} 
	\centering
		\includegraphics[width=1.00\textwidth]{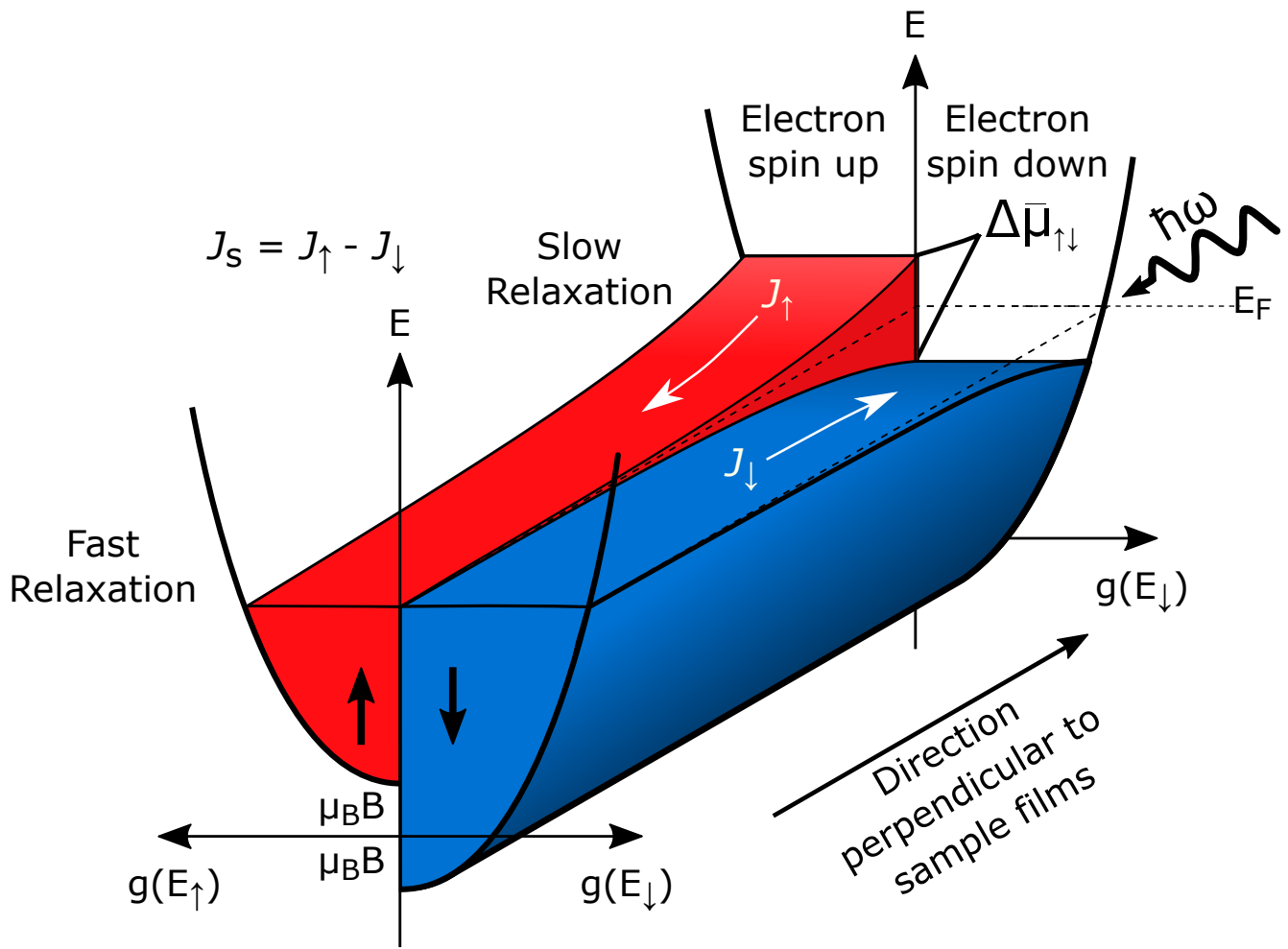}
	\caption{Spin channels in an Overhauser-pumped material with a short-circuiting interface. The Electrochemical potential gradients established in each spin channel due to the differing relaxation rates of the adjacent layers generate the ``impure'' spin currents $J_{\uparrow}$ and $J_{\downarrow}$. Considering both of these flows simultaneously allows us to observe a net flow of angular momentum in the absence of a net flow of charge - a ``pure'' spin current $\Js$.}
	\label{fig:SpinBattery}
\end{figure}

The Overhauser-induced spin accumulation of Figure \ref{fig:PauliDiagram}d may be turned into a simple microwave Spin Battery which generates pure spin currents that are unencumbered by charge transport. This is achieved by engineering a gradient of spin accumulation and this is in turn arranged by making provision to ``short-circuit'' the two spin channels by introducing rapid spin-flipping - thereby reducing the electrochemical splitting to negligible magnitude - on one side of the microwave irradiated material. The spin channel potentials then assume the form shown in Figure \ref{fig:SpinBattery}.

The form of these spin potentials bears some resemblance to those found at a ferromagnetic/paramagnetic interface in Current-Perpendicular-to-Plane Giant Magnetoresistance (CPP GMR)\cite{Valet1993} with the difference that, although $\dmud$ has a negative gradient in both cases, the individual spin channel gradients have the same sign in the CPP GMR case and opposite signs in our Spin Battery as a reflection of the fact that the latter case represents pure spin current uncontaminated by charge current.

\begin{figure}
	\centering
		\includegraphics[width=1.00\textwidth]{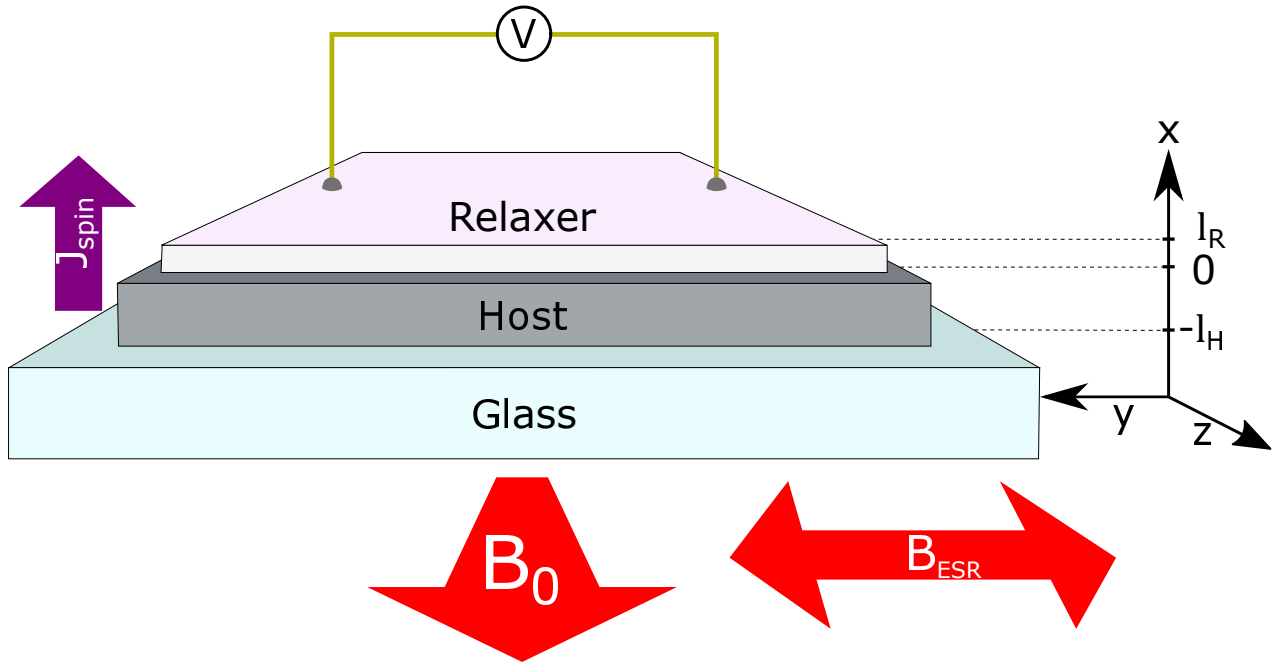}
	\caption{A thin film Spin Battery device. The device has been set up to make Inverse Spin Hall Effect (ISHE)\cite{doi:10.1063/1.2199473} measurements assuming the relaxer has a large Spin Hall Angle.}
	\label{fig:Setup}
\end{figure}

The short-circuiting of the spin channel potentials may be achieved by bringing the Overhauser host metal into electrical contact with another conducting material that has a faster spin-flip relaxation rate. A practical embodiment of such a device is shown in Figure \ref{fig:Setup} which depicts a cross-section through a thin film bilayer. A similar outcome could be achieved by only irradiating half of the host material, as in \cite{Watts2006}, however this is experimentally challenging.

The darker layer is the Overhauser host layer that comprises either a ferromagnetic metal as discussed above, or a paramagnetic metal with a long relaxation time such as pure silver, which has a spin diffusion length of about a micron\cite{Kimura2007}. This layer is microwave-irradiated at its magnetic resonance frequency to generate an Overhauser-induced spin accumulation. The contiguous lighter layer is a paramagnetic metal with a short spin relaxation time - for example platinum; or a silver alloy containing a few percent of gold which drastically reduces the spin diffusion length to the order of 10nm by introducing spin-orbit scattering \cite{Gregg1997}.

The sample dimensions are small compared with the electromagnetic skin depth \cite{jackson2007classical} so that Overhauser spin pumping may occur throughout the sample. Moreover, given Fermi velocities of order $10^{6}$ metres per second and device thicknesses of order nanometres, conduction electrons in such thin film metallic bilayers sample both layers on a femtosecond timescale which is four orders of magnitude shorter than the tens-of-picoseconds timescale of the ESR Larmor frequency. Silsbee et al. \cite{PhysRevB.19.4382} examines the effect on the ESR signal of this temporal imbalance. The ESR linewidth, strength and effective g-value are a function of both materials in the bilayer and the resonance is averaged in a not dissimilar way to the operation of motional narrowing in liquid NMR. It follows on from these considerations that in deriving a properly rigorous model for the spatial variation of the microwave-induced spin accumulation, it is necessary to make a fully quantum-statistical analysis that takes these very different timescales into account. Nevertheless, the metal with the faster spin relaxation generates a smaller spin accumulation with a shorter decay length (i.e. a smaller spin diffusion length), so when the metals are in contact this gives rise to varying spin channel electrochemical potentials and hence spin accumulation gradients occur in both metals.

In the early days of Giant Magnetoresistance it was tacitly assumed that all spin transport was by virtue of electrical carrier displacement, and, for simplicity, this is the assumption we have made for the bilayer interface. Moreover, as it is a metal-metal interface, it is essentially free of the impedance mismatch issues that concern Ando et al. \cite{Ando2011}. However, modern thinking \cite{PhysRevB.89.174417, PhysRevLett.116.077203} suggests that mutual spin-flip processes may also play a significant role in spin transport across interfaces \cite{BASS2016244,doi:10.1063/1.4989678} and any detailed quantitative model needs to consider this additional possibility.

The geometry of our proposed two layer Spin Battery device was conceived, designed and its dimensions optimised \textbf{entirely} in the light of the Overhauser science that we discuss above and the Valet-Fert model that we invoke to model the spin current yield. We now compare the design that has emerged with the large family of microwave spin pumping devices in the contemporary literature and, in the language of patent attorneys, we find that it passes The Duck Test (``if it walks like a duck, swims like a duck and quacks like a duck then it \textbf{IS} a duck''). Serendipitously, the structure that has emerged from our Overhauser-driven design considerations \textbf{looks and behaves exactly like} all recently reported Spintronic Microwave Spin Pumping devices \cite{doi:10.1063/1.4871514,PhysRevB.89.235317,doi:10.1063/1.4792693,doi:10.1063/1.3587173,doi:10.1063/1.4918909,PhysRevB.82.214403,Ando2011}: and is essentially a distillation of their common denominators: specifically an Overhauser pumping zone where a spin accumulation is formed and a symmetry-breaking element that causes a spin accumulation gradient that in turn gives rise to the desired spin current. This encourages us to believe that the Overhauser pumping scheme we have described is indeed the basic underlying principle of contemporary Spintronic Microwave Spin Pumping.

Unmasking the Overhauser Effect as the ``secret agent'' responsible for contemporary spin pumping leads to some intriguing predictions for further experiments. For example, experimental observations of spin pumping to date have all been in magnetic materials: but the Overhauser Effect is equally applicable to paramagnetic conductors so that Spintronic Microwave Spin Pumping should be observable in those also in applied magnetic fields that are sufficiently large to afford adequate signal to noise. It is interesting to note that, since Carver and Slichter \cite{Carver1956} were able to detect spin pumping in very small magnetic fields (of order a few milliTesla) in (non-magnetic) Lithium metal by measuring the resulting nuclear polarisation, this implies that NMR is experimentally a more sensitive probe of the existence of a spin accumulation than is the Inverse Spin Hall Effect (ISHE).

Moreover, the sophisticated toolkit of the world of magnetic resonance comprises numerous high sensitivity double resonance techniques such as Electron-Nuclear Double Resonance (ENDOR) by which the effect of changing the spin states of the nuclei may be detected by their effect on the electronic resonance. Analogous effects should therefore be observable in spin pumping. We have been privileged to see early experimental results in that vein that are the subject of a forthcoming publication \cite{SaitohUnpublished}.

In conclusion, we have proposed a spin pumping mechanism based on the longstanding Overhauser Effect which we believe is the microscopic mechanism that underpins recent observations of spin pumping in metallic spintronic systems.  We note that our proposed mechanism functions by creating what is known in the modern field of Spintronics as a spin accumulation. Recent observations by the semiconductor community of spin-accumulation-induced Dynamic Nuclear Polarisation \cite{Shiogai2012} reinforces this observation. We have made and tested a simple, two-layer, unlithographed Spin Battery device that works on the Overhauser principle and that, although designed exclusively with the Overhauser Effect in mind, serendipitously incorporates a distillation of the essential features of most of the spintronic spin-pumping appliances already reported. This strengthens our belief that the mechanism that we propose has already been multiply observed in the existing literature but that its intimate connection with $20^{\textrm{th}}$ Century work in the field of Magnetic Resonance has not been recognised. We predict that in sufficiently large applied magnetic fields the Overhauser Effect can provide for microwave spin-current generation in magnetic and non-magnetic materials alike. We note that NMR is potentially a more sensitive experimental diagnostic tool than ISHE. We also predict that there is scope for multiple double resonance spin pumping effects that are analogous to magnetic resonance phenomena such as ENDOR and NEDOR.

\section*{Acknowledgments}The authors thank James Binney for very helpful discussions, Arzhang Ardavan and Junjie Liu for ESR measurements and Eiji Saitoh for sharing experimental results prior to publication. AI and CJT respectively thank Magdalen College for a Beghian studentship and EPSRC for doctoral funding.
\section*{Author Contributions}JFG recognised the potential for exploiting the Overhauser effect in Spintronics. CJT, JFG, and BH developed the theoretical model for this work, samples were grown by CJT, EP, and MEM, measurements were made by CJT, AI, and MEM, JFG, BH, AI, MEM, EP, JS, and CJT participated in the discussions, and JFG, CJT, and JS prepared the manuscript.
\section*{Competing Interests} The authors declare no competing financial interests.

\bibliography{Overhauserpaperbib.bib}{}

\end{document}